\documentclass[aps, groupedaddress,twocolumn,amsmath,amssymb]{revtex4}
\usepackage{braket}
\usepackage{mathptm}
\usepackage{times}
\usepackage{amsmath}
\usepackage{dcolumn}
\usepackage{graphicx}
\usepackage{color}
\usepackage{ulem}
\usepackage{subfigure}

\begin{document}

\title{A flexible multi-reference perturbation theory by minimizing the Hylleraas functional with matrix product states}

\author{Sandeep Sharma}
\affiliation{Department of Chemistry, Frick Laboratory, Princeton University, NJ 08544}
\author{Garnet Kin-Lic Chan}
\affiliation{Department of Chemistry, Frick Laboratory, Princeton University, NJ 08544}
\email{gkc1000@gmail.com}

\begin{abstract}
We describe a formulation of multi-reference perturbation theory that obtains a rigorous upper bound to the second order energy by minimizing the Hylleraas functional 
in the space of matrix product states (MPS). The first order wavefunctions so obtained 
can also be used to compute the third order energy with little overhead.
Our formulation has several advantages including (i) flexibility with respect to the choice of zeroth order Hamiltonian, 
(ii) recovery of the exact uncontracted multi-reference
perturbation theory energies in the limit of large MPS bond dimension, (iii) no 
requirement to compute high body density matrices, 
(iv) an embarrassingly parallel algorithm (scaling up to the  number of virtual orbitals, squared, processors). Preliminary numerical examples 
show that the MPS bond dimension required for accurate first order wavefunctions scales sub-linearly with the size of the basis. 
\end{abstract}

\maketitle

%\section{Introduction}
Commonly, electron correlation is classified into static (or strong) correlation and dynamic (or weak) correlation. The former is essential to capture the qualitative
electronic structure, and the latter to generate quantitatively accurate results. In most chemical systems,  static correlation involves a small subset of orbitals 
with degenerate orbital energies on the energy scale of the Coulomb repulsion. In this subspace, also known as the active space, multi-configuration self-consistent field (MCSCF) 
calculations have traditionally been performed. However, the cost of (numerically exact) MCSCF scales exponentially with the number of active orbitals and is thus 
 restricted to active spaces with about 16 electrons in 16 orbitals. In the last decade, this active space restriction has effectively been removed, without  significant
numerical errors, with the advent of new near-exact methods such as the density matrix renormalization group (DMRG)~\cite{whiteqm,chan2011,legeza-rev,marti11,yuki2,wouters-review,Kurashige2013, sharma2014} and full configuration interaction quantum Monte Carlo (FCIQMC)~\cite{booth,booth1,Booth-neci}.
(More approximate techniques for large active spaces, such as restricted active space and general active space methods~\cite{rasscf,gasscf}, high-order active-space coupled 
cluster~\cite{kallay,ccvb,piecuch1999coupled,olsen-cc}, and variational reduced density matrix methods~\cite{nakatsuji,mazziotti,brecht-pra}, have also been advanced). 
However, for quantitative chemical accuracy, these new active space methods must still be augmented with techniques to include dynamic correlation, and this remains an important frontier~\cite{yuki,be2sandeep,dmrgct,dmrgmrci,gag-gaspt2}.

The primary technical difficulty in treating dynamic correlation is the large
number of orbitals require to converge the short-range electron-electron cusp. Explicit correlation (through R12 and F12 factors)~\cite{tenno-rev, torheyden,Kong2011,klopper,yanai2012canonical,manby} significantly amelioriates,
but does not eliminate, the need for large basis sets. 
One affordable approach to include dynamic correlation from a large number of external orbitals is through perturbation theory.
Examples of such multi-reference perturbation methods include CASPT2\cite{caspt2,caspt2-rev} and NEVPT2\cite{nevpt2,nevpt2-rev}. However, even when recently combined with 
DMRG reference functions\cite{yuki,dmrgpt22}, these methods have only been used with 
moderate active space sizes. The key bottleneck is the need to
construct intermediates involving three- and four-body reduced density matrices (RDMs) in the active space.  As the number of active orbitals increases,
calculating and storing these RDMs becomes prohibitively expensive. 

In this communication, we describe an alternate formulation of a multi-reference perturbation theory, which we call matrix product state perturbation theory (MPS-PT). The new formulation possesses substantial advantages.  
\begin{enumerate}
\item It bypasses the need for high-body RDMs in the active space, potentially allowing very large
active spaces to be used.
 \item Perturbation theory depends strongly on the zeroth-order Hamiltonian $H_0$. 
Changing $H_0$ (as in CASPT2 versus NEVPT2) usually requires re-deriving and re-implementing non-trivial intermediates. In our formulation we 
can easily change $H_0$ without significantly changing the implementation.
 \item To reduce the cost of multi-reference perturbation theory, the first order wavefunction is 
often determined in a restricted space, as in internally contracted CASPT2 or  partially and strongly contracted NEVPT2 (PC-NEVPT2, SC-NEVPT2). Aggressive contraction 
can lead to additional errors which are difficult to estimate \textit{a priori}. We also 
contract the first order space via the MPS bond dimension,  but do so in an automatic and optimal 
way. This contraction systematically and rapidly
converges to the uncontracted result with increasing bond dimension, and the errors of 
contraction can be robustly estimated.
\item The algorithm is highly parallel and has a modest scaling with the 
MPS bond dimension and size of basis.
\end{enumerate}

It is well known that perturbation theory can be formulated as a variational problem\cite{oktay, hylleraas}. For the second order energy, the variational functional is the Hylleraas functional $H[\Psi_1]$, Eq.~(\ref{eq:hyll}), which is minimized with respect to the first order wavefunction $|\Psi_1\rangle$,
\begin{align}
 H[\Psi_1] = \langle\Psi_1| H_0 - E_0|\Psi_1\rangle + 2 \langle\Psi_1|QV|\Psi_0\rangle. \label{eq:hyll}
\end{align}
Here $H_0$, $E_0$ and $|\Psi_0\rangle$ are respectively the zeroth order Hamiltonian, energy and wavefunction, $V$ is the perturbation ($H-H_0$), and $Q$  the projector onto the space 
orthogonal to $|\Psi_0\rangle$.   
It can be easily verified that at the minimum, the wavefunction $|\Psi_1\rangle$ satisfies the familiar equation
\begin{align}
 (H_0 - E_0)|\Psi_1\rangle = -QV|\Psi_0\rangle \label{eq:pt2}.
\end{align}

\begin{figure*}
\begin{center}
\includegraphics[width=0.98\textwidth]{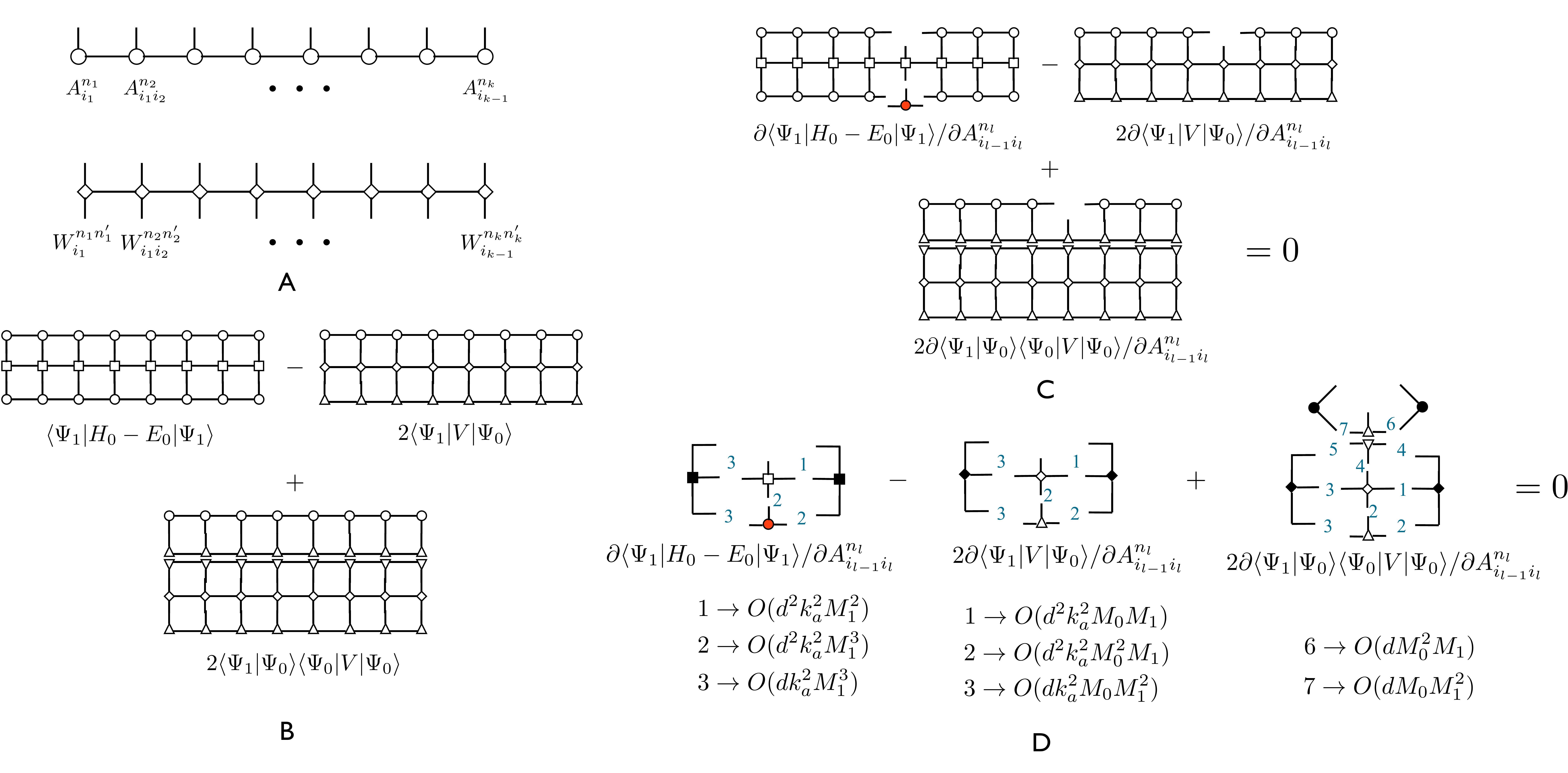}
\end{center}
\caption{[\textbf{A}] A matrix product state (MPS) can be represented graphically using a series of 3-dimensional tensors, shown with circles, each of which is associated with an orbital. 
The free index, also known as the physical index (pointing upwards) of the tensors denotes the occupation of the orbital and the other two indices, known as auxiliary indices, are 
sequentially contracted. The dimension $d$ of the physical index is 4 for a spatial orbital, and the dimension $M$ of the auxiliary index can be increased to make the MPS arbitrarily flexible. 
Similarly, a matrix product operator (MPO) can be represented as a series of 4-dimensional tensors, with two physical indices and auxiliary indices contracted sequentially. 
Due to the 2-body nature of the Hamiltonians in quantum chemistry, the auxiliary bond dimension of an MPO is always less than $k^2$, where $k$ is the number of orbitals. [\textbf{B} and \textbf{C}] The tensors in wavefunctions $\Psi_1$ and $\Psi_0$ are represented by circles and triangles respectively. The zeroth order Hamiltonian $H_0$ and perturbation operator $V$  
are represented by squares and diamonds respectively. Panel B shows the three terms in the Hylleraas functional. Panel C shows the partial derivative of the Hylleraas functional $\partial H[\Psi_1]/\partial A^{n_l}_{i_{l-1}i_l}$ that is set to zero in the optimization. The red tensor, $A^{n_l}_{i_{l-1}i_l}$ is the quantity being solved for. [\textbf{D}] Contractions that need to be carried out at each step of the sweep iteration. 
The corresponding costs (assuming $H_0$ is Dyall's Hamiltonian\cite{dyall}) are shown. For each term, individual contractions are numbered according to the order in which they are performed. 
If two contractions have the same number, it means that the two indices on the same tensor are fused to form a larger composite index and then the contraction is performed. 
The solid tensors represent the results of contracting all the open tensors in panel C that are not shown in this figure.}
\label{fig:mps}
\end{figure*}

In MPS-PT2 we minimize the Hylleraas functional, while expressing $|\Psi_1\rangle$ as a matrix product state (MPS). MPS form a complete variational space,
and the quality of the MPS solution is controlled by a single parameter ``$M$'',  the dimension of the auxiliary bond in the MPS (see Figure~\ref{fig:mps}).
Overlaps ($\langle\Psi_1|\Psi_2\rangle$) and transition elements of operators between two MPSs ($\langle\Psi_1|O|\Psi_2\rangle$) can be evaluated in $O(M^3)$ CPU time. The Hylleraas functional can be minimized with respect to an MPS of arbitrary bond dimension using a sweep algorithm analogous to that in the density matrix renormalization group (DMRG)\cite{chan2002}.
In the limit of large $M$, the solution converges to the exact (uncontracted) first order 
wavefunction (and second order energy).
From the first order wavefunction, the third order correction can also be calculated due to Wigner's $2n+1$ rule, as
\begin{align}
 E_3 = \langle \Psi_1|V|\Psi_1\rangle - E_1 \langle \Psi_1|\Psi_1\rangle. \label{eq:3rd}
\end{align}

An MPS for $|\Psi_1\rangle$  is shown in Eq.~(\ref{eq:mps1}), where $n_i$ is the occupation of orbital $i$,
\begin{align}
 |\Psi_1\rangle = \sum_{\{n\}} A^{n_1}_{i_1} A^{n_2}_{i_1 i_2} \ldots A^{n_k}_{i_{k-1}} |n_1 n_2 \ldots n_k\rangle. \label{eq:mps1}
\end{align}
Eq.~(\ref{eq:mps1}) is conveniently expressed graphically as a tensor network, as
shown in panel A of Figure~\ref{fig:mps} (see article~\cite{Schollwock2011,chan-wiley,Orus} for a detailed explanation of the 
graphical notation). Similarly, any operator $\Omega$ can be written in matrix product operator (MPO) form,
\begin{align}
 \Omega = \sum_{\{n\}} W^{n_1 n_1'}_{i_1} W^{n_2 n_2'}_{i_1 i_2} \ldots W^{n_k n_k'}_{i_{k-1}} |n_1 n_2 \ldots n_k\rangle \langle n_1 n_2 \ldots n_k|\label{eq:mpo}
\end{align}
and this is also expressed graphically in panel A of Figure~\ref{fig:mps}.
With this  notation, the Hylleraas functional is shown in panel B of Figure~\ref{fig:mps}, where the operators and wavefunctions
have been represented as MPOs and MPSs respectively, and the corresponding tensor networks are contracted to obtain the final expression.

In the sweep algorithm, each tensor $[A^{n}]$ of the MPS wavefunction $|\Psi_1\rangle$ is optimized in sequence, keeping the other tensors fixed. This converts the non-linear 
optimization of the MPS (which is multi-linear in its parameters) to a series of linear equations that we solve using the conjugate gradient algorithm. 
At step $l$ of the sweep, we take the partial derivative of the Hylleraas functional with respect to the set of tensor elements $A^{n_l}_{i_{l-1} i_l}$ of $|\Psi_1\rangle$, and set it to zero.
The resulting linear equation is shown graphically in panel C of Figure~\ref{fig:mps} and has 3 terms, corresponding to the partial derivatives of the three terms in the Hylleraas functional. 
The first involves a contraction of $A^{n_l}_{i_{l-1} i_l}$ (the quantity being solved for, shown with a red circle) 
with a 6-index tensor, formed by contracting all tensors in the operator $H_0-E_0$ and  $|\Psi_1\rangle$ with
 $A^{n_l}_{i_{l-1} i_l}$  removed. Similarly, the second and third terms involve the tensors of $V$, $|\Psi_0\rangle$ and all tensors of $|\Psi_1\rangle$ with $A^{n_l}_{i_{l-1} i_l}$
 removed. The resulting linear equation for $A^{n_l}_{i_{l-1} i_l}$ is in $O(dM^2)$ unknowns. Solving it by conjugate gradient requires only $O(M^3)$ time (due to the
special structure of the operator-vector product) and yields the current best estimate 
of the tensor $A^{n_l}_{i_{l-1} i_l}$. It is clear that the tensor so obtained depends on the values of all the other tensors $[A^{n}]$ in  $|\Psi_1\rangle$. Thus 
we sweep through all the orbitals of $|\Psi_1\rangle$ solving a linear algebra problem at
each step $l=1\ldots k$ to optimize each tensor, while keeping the others fixed. 
This algorithm is essentially  the same as the standard DMRG sweep algorithm,  the only difference being that instead of an eigenvalue problem, a linear equation is solved at each step.

%\subsection*{Computational costs}%parallelization, symmetries
The computational cost of the above sweep can be expressed in detail in terms of the auxiliary bond dimensions  of $|\Psi_1\rangle$ and $|\Psi_0\rangle$ ($M_1$ and $M_0$, respectively) and the number of active and external orbitals ($k_a$ and $k_v$, respectively).
%\begin{itemize}
% \item 
For concreteness, we consider $H_0$ to be Dyall's Hamiltonian\cite{dyall}, as used in NEVPT2. 
All the contractions  to be performed at each sweep iteration, together with their respective (leading order in $M$, $k_a$, $k_v$) computational
costs, are shown in Panel D of Figure~\ref{fig:mps}. 
Contractions 1, 2 and 3 of the first two terms have the same leading scaling with respect to the number of orbitals
even though the operator in the first term (denoted by squares) is a Dyall Hamiltonian $H_0$, which does not couple the active and external spaces, and the operator $V$ in the 
second term (denoted by diamonds) couples the two spaces. The  low scaling of the second term is because components of $V$ with more than 2 external indices do not contribute to the 
second order energy and can be neglected. The resulting MPO for $V$, under this restriction, can then always be represented with  $O(k_a^2)$ auxiliary bond dimension, the same as the Dyall $H_0$. The costs for contractions 1 to 5 in the third term are not shown because they are used to construct the bottom diagram, which is a constant that does not change during the sweep, and thus
has a subleading dependence on $k_v$. The final computational cost of a single sweep is $O(k_vk_a^2M_1^3 + k_vk_a^2M_0M_1^2 + k_vk_a^2M_0^2M_1)$, where we have multiplied the most expensive 
costs in each individual step shown in Panel D of Figure~\ref{fig:mps} by $k_v$ (the number of virtual orbitals) to reflect the total number of steps in the sweep, assuming $k_v+k_a \sim O(k_v)$.
We have also dropped the dependence on $d$, which is a fixed constant (usually 4). 
(For brevity, we have
not considered the cost of terms with 
subleading (lower than cubic) dependence on $M$. These only become important
when $k_v$ is very large. For example,  forming the MPO for $V$ (through DMRG complementary operator techniques~\cite{xiang}) costs $O(k_v^2k_a^2M_0M_1)$ per sweep, which becomes important when $k_v>M$). 

The above costs  assume the Dyall $H_0$. A similar analysis applies to the CASPT2 $H_0$, which takes the form
of a one-body Fock operator multipled by many-body projectors. The many-body projectors lead to additional 
overlap computations of the form $\langle \Psi_0|\Psi_1 \rangle$, but these only cost $O(k_aM_0^2M_1 + k_aM_0M_1^2)$ for the entire sweep, which is independent of $k_v$. The one-body Fock operator
leads to a lower overall computational cost for the first term in
the Hylleraas functional involving  $H_0$. However, the second term involving $V$ results in the same leading cost as for NEVPT2 above. Additionally, 
note that because we do not proceed via special reduced density matrix intermediates,
the computations for both the Dyall $H_0$ in NEVPT and for the CASPT2 $H_0$ involve the
same basic contractions of MPOs and MPSs
and thus it is simple to change the zeroth order Hamiltonian in the implementation.
This is similar to the simplicity of implementing fully uncontracted determinant algorithms, except
here contraction is automatically provided by the finite bond dimension of the MPS.

%\subsection*{Implementation}
We have implemented the above MPS-PT sweep algorithm to evaluate the NEVPT2 and NEVPT3 energy in the \textsc{Block} code\cite{sharma2012spin}. For computational
robustness, we have implemented the two-site version of the sweep algorithm (see Refs.~\cite{chan2002,Schollwock2011} for a discussion of the difference between one-site
and two-site sweep algorithms).
In the limit of large $M_1$, our MPS-PT algorithm will yield the uncontracted NEVPT2 and NEVPT3 energies.
We have parallelized the algorithm exactly 
in the manner of the DMRG module of the \textsc{Block} code, where each auxiliary
bond contributing to the MPO (arising from a term in the operator/complementary operator product) is evaluated on a different processor. As explained previously, the auxiliary bond dimension of 
both $H_0$ and $V$ is $O(k_a^2)$, thus the code is expected to efficiently parallelize over $O(k_a^2)$ processors. We refer the reader to the detailed description of DMRG parallelization in reference \cite{chan2004}. There is an additional ``embarrassing'' parallelization over $O(k_v^2)$ processors which we mention (but have not implemented). This arises by breaking the 
summation in $V$ into  $k_v^2$ 
individual terms $V_i$. Each $V_i$ produces a mutually orthogonal state when  acting on $|\Psi_0\rangle$ and thus  the Hylleraas functional can be computed and minimized for 
each $V_i$ separately, giving a final second-order energy that is just the sum of these terms. 
 
Symmetries of the Hamiltonian are easily utilized in an analogous fashion to the DMRG module of the \textsc{Block} code, which currently  implements many symmetries including
particle number, Abelian and non-Abelian point groups, and SU(2) (spin symmetry).

\begin{table}
\caption{Energy (E+108.0 in E$_h$) for an MPS-PT calculation performed with a cc-pVDZ basis set for the $^1\Sigma_g$ state of N$_2$ molecule at various bond-lengths. A (10e, 8o) active space was used. PC-NEVPT2, and SC-NEVPT2 results are given for comparison. 
The last four columns show the MPS-PT2 energy calculated using different auxiliary bond dimensions $M_1$.}\label{tab:n21}
\begin{center}
\begin{tabular} {lccccccc}
 \hline
 \hline
 $r_e$ & PC- & SC- &MPS-PT3&\multicolumn{4}{c}{$M_1$ used in MPS-PT2}\\
 (\AA) &NEVPT2&NEVPT2&M=800&800&300&200&100\\
 \cline{1-4}\cline{5-8}
1.8		&	-1.1430	&	-1.1396	&	-1.1626	&	-1.1432	&	-1.1431	&	-1.1428	&	-1.1403	\\
2.0		&	-1.2420	&	-1.2389	&	-1.2620	&	-1.2422	&	-1.2421	&	-1.2418	&	-1.2382	\\
2.2		&	-1.2487	&	-1.2458	&	-1.2689	&	-1.2490	&	-1.2487	&	-1.2483	&	-1.2432	\\
2.4		&	-1.2129	&	-1.2103	&	-1.2333	&	-1.2132	&	-1.2129	&	-1.2123	&	-1.2066	\\
2.7		&	-1.1342	&	-1.1318	&	-1.1549	&	-1.1345	&	-1.1341	&	-1.1332	&	-1.1258	\\
3.0		&	-1.0591	&	-1.0568	&	-1.0807	&	-1.0594	&	-1.0589	&	-1.0578	&	-1.0484	\\
3.6		&	-0.9648	&	-0.9631	&	-0.9896	&	-0.9651	&	-0.9641	&	-0.9624	&	-0.9430	\\
4.2		&	-0.9350	&	-0.9342	&	-0.9621	&	-0.9352	&	-0.9341	&	-0.9324	&	-0.9055	\\
\hline
\end{tabular}
\end{center}
\end{table}

\begin{figure}
\begin{center}
\includegraphics[width=0.4\textwidth]{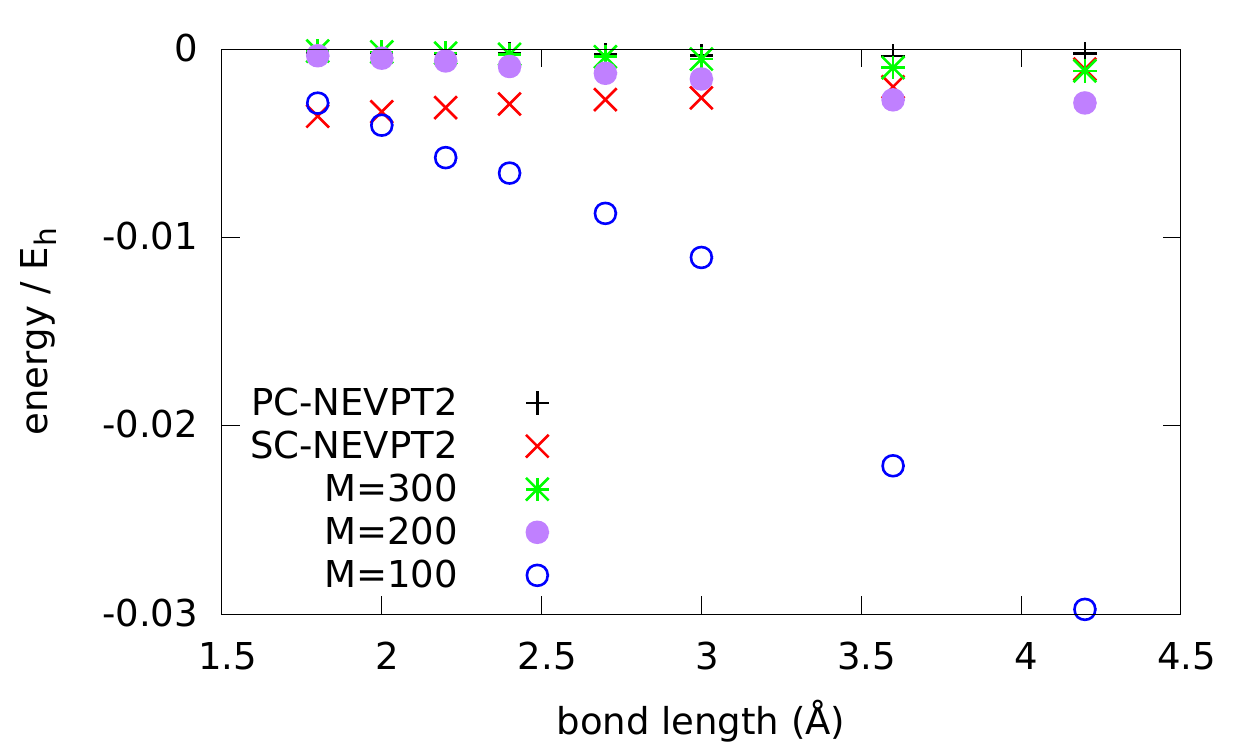}
\end{center}
\caption{Error in the second-order energy (in E$_h$) for various flavors of NEVPT2 and MPS-PT2 calculations performed with cc-pVDZ basis sets for the $^1\Sigma_g$ state of N$_2$ molecule at several bond lengths. A (10e, 8o) active space was used for these calculations. The graph shows that the value of $M_1$ required 
to capture the 
correct second-order energy using the MPS-PT2 theory increases with the N$_2$ bond length, most likely due to increase in the multi-reference character of the wavefunction. Nonetheless,
for $M_1>100$, the error remains quite small at all bond-lengths. }
\label{fig:n2diss}
\end{figure}

%
%\section{Nitrogen dissociation curve}
We next illustrate various aspects of the theory by performing calculations on the N$_2$ bond dissociation benchmark using a full-valence CAS of 10 electrons
in 8 orbitals and the cc-pVDZ, cc-pVTZ, and cc-pVQZ basis sets~\cite{dunningbasis}.
Table~\ref{tab:n21} shows the energies of the N$_2$ molecule with various NEVPT2 and MPS-PT variants at several bond-lengths in the cc-pVDZ basis set. The MPS-PT2 energy is 
converged to a few tens of $\mu$E$_h$ with $M_1$=800 and always gives an energy lower than the partially contracted NEVPT2 theory (since it is converging to the
fully uncontracted result) although the energy difference is not large. The error of various standard flavors of contracted NEVPT2 theory, PC-NEVPT2 and SC-NEVPT2 (evaluated
using \textsc{Molpro}\cite{werner2012} and MPS-PT2 calculations with smaller $M_1$=100-300, are plotted against increasing N$_2$ bond-length in Figure~\ref{fig:n2diss}. 
We see that the value of $M_1$ required to accurately describe the first order wavefunction $|\Psi_1\rangle$  increases somewhat with increasing bond length. 
This can be understood by realising that the first-order wavefunction contains contributions from $O(N_a k_a^2k_v^2)$ determinants (where $N_a$ is the number
of significant determinants in the active space). As $N_a$ increases at increased bond-lengths, the first-order wavefunction becomes more complicated and requires slightly larger $M$. Nonetheless, 
$M\ll N_a k_a^2 k_v^2$ and except for $M=100$, the MPS-PT error remains  small at all bond-lengths.

\begin{figure}
\begin{center}
\includegraphics[width=0.4\textwidth]{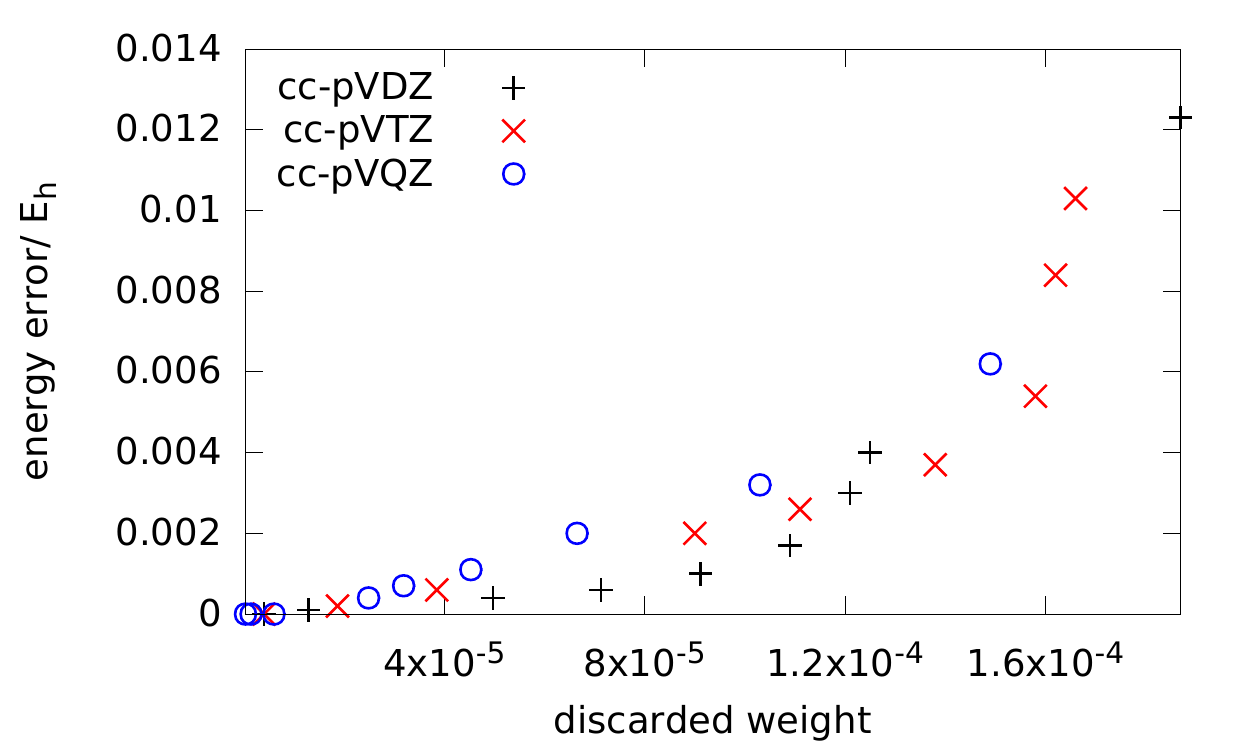}
\end{center}
\caption{Error in the second-order energy (in E$_h$) versus the discarded weight for a MPS-PT2 calculation performed with cc-pVDZ, cc-pVTZ, and cc-pVQZ basis sets for the $^1\Sigma_g$ state of N$_2$ molecule at a bond length of 2~\AA. A (10e, 8o) active space was used for these calculations. After an initial rapid relaxation of the energy one recovers the linear relationship between the energy and the discarded weight.}
\label{fig:discard}
\end{figure}

When one views the DMRG algorithm in the traditional renormalization group language, each tensor in the MPS (in canonical form) can be viewed as mapping from $4M$ states (for $d=4$) to the $M$ most 
significant renormalized states. In the two-site variant of the algorithm, the weight of these discarded $3M$ states measures the severity of the truncation performed in 
the renormalization step. The discarded weight goes to zero in the limit of  large $M$ and for smaller  $M$  usually exhibits a 
linear relationship with the energy error~\cite{legeza,chan2002}. Figure~\ref{fig:discard} shows that after an initial rapid relaxation in the energy, the error in the MPS-PT2 indeed scales linearly with the 
corresponding discarded weight in $\ket{\Psi_1}$. Thus the relationship between the energy error and discarded weight can be used to gauge the convergence of the MPS-PT2 energy with $M$,
 just as in the DMRG algorithm.

\begin{figure}
\begin{center}
\includegraphics[width=0.4\textwidth]{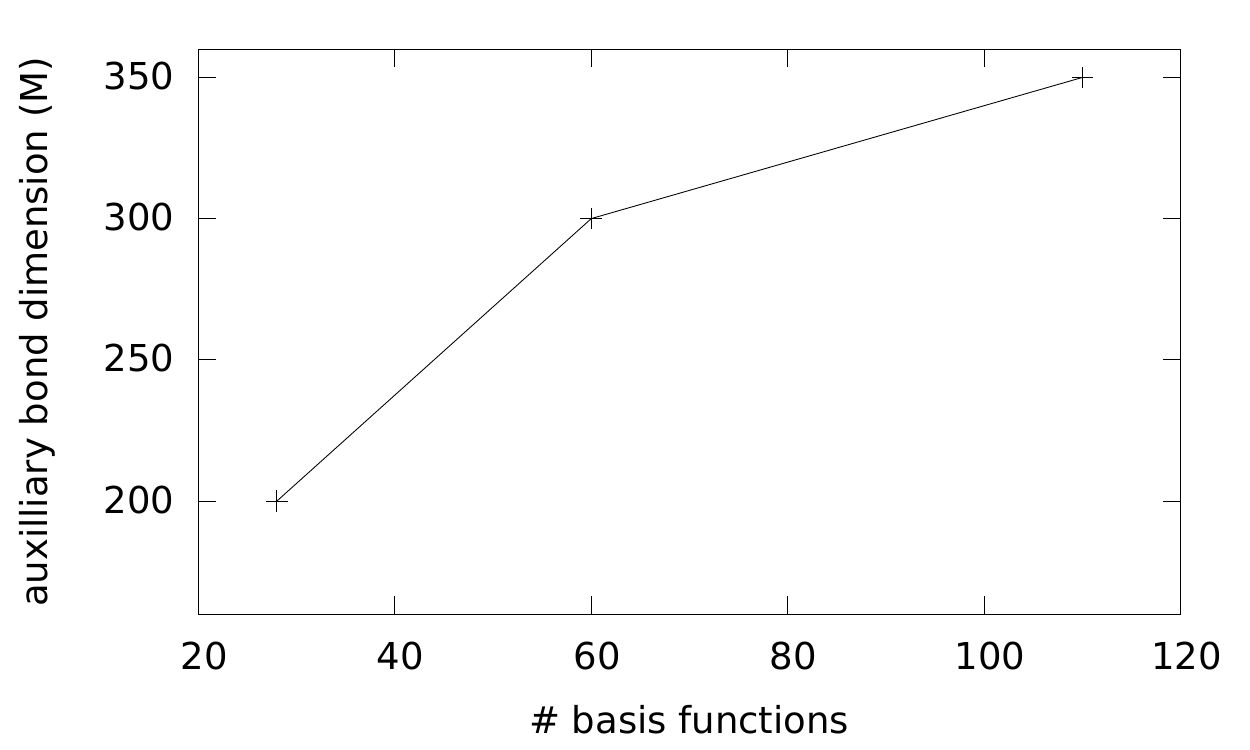}
\end{center}
\caption{The size of the auxiliary bond dimension $M$ of the first order wavefunction required to obtain a second order energy within 1 mE$_h$ of the exact fully uncontracted MPS-PT2 energy.}
\label{fig:Mscale}
\end{figure}

Figure~\ref{fig:Mscale} shows the relationship between the size of the basis set and the value of $M_1$ required to obtain a second-order energy within 1 mE$_h$ of the exact 
fully uncontracted MPS-PT2 energy. We see that, encouragingly, the $M_1$ required scales sub-linearly with the size of the basis. Indeed, the entanglement in the basis set
limit is bounded by a constant (the dimension of the active orbital Hilbert space). Thus, $M_1$ must asymptotically be independent of $k_v$.

%% The graph is obtained by performing MPS-PT2 calculations at each basis set with progressively increasing M in steps of 50 until we reach an error in the second order energy of lower than 1 mE$_h$. The graph shows that the size $M$ required scales somewhat more favourably that linearly with the size of the basis set. 

To summarize, we have shown that minimizing the Hylleraas functional in the space of wavefunctions described by an MPS can be used to obtain accurate upper bounds to the exact multi-reference
second order perturbation theory energy. With little additional computational overhead, third order corrections to the energies can also be obtained. No active space density matrices are required, allowing
large active spaces to be treated, and our implementation can be easily modified to accommodate
 any reasonable zeroth order hamiltonian. Parallelism is also easily exploited.
In benchmark calculations we have shown that this approach efficiently recovers the uncontracted perturbation theory
energy, beyond existing internally contracted multireference perturbation results. Finally, we 
observe  that the bond dimension of the first order wavefunction required to obtain chemical accuracy scales sub-linearly with the size of the basis set, which is promising for future applications
with very large basis sets.

\begin{acknowledgements}
%\textbf{Acknowledgements}\\
This work was supported by the US National Science Foundation (NSF) through Grant No. 
NSF-CHE-1265277. Additional support for software development was provided
through Grant NSF-OCI-1265278. S. S. would also like to thank Brecht Verstichel for many helpful discussions.
\end{acknowledgements}

%\bibliographystyle{Science}
%\bibliographystyle{apsrev4-1}
%\bibliography{references}

\end{document}